\documentclass[aps,prd,preprint,tightenlines,superscriptaddress,
amsfonts,amssymb,amsmath,showpacs,nofootinbib]{revtex4-1}
\usepackage{graphicx}
\usepackage{amssymb,amsmath,amsfonts,palatino,amsthm}
\usepackage{amssymb}
\usepackage{epstopdf}
\usepackage{color}

\newcommand{\Bra}[1]{\langle #1 \vert} 
\newcommand{\Ket}[1]{\vert #1 \rangle} 
\newcommand{\BraKet}[2]{\langle #1 \vert #2 \rangle} 

\begin{document}

\newcommand{\dR}{\mathbb R}
\newcommand{\dC}{\mathbb C}
\newcommand{\dZ}{\mathbb Z}
\newcommand{\id}{\mathbb I}
\newcommand{\dT}{\mathbb T}
\newcommand{\UnitOp}{\hat{\mathbb{I}}} 
\newcommand{\Group}[1]{\mathrm{#1}} 
\newcommand{\Aver}[1]{\langle #1 \rangle}

\title{Generic singularity of general relativity and its quantum fate}

\author{W{\l}odzimierz Piechocki} \email{wlodzimierz.piechocki@ncbj.gov.pl}
\affiliation{Department of Fundamental Research, National Centre for Nuclear
  Research, Pasteura 7, 02-093 Warsaw, Poland}

\date{\today}

\begin{abstract}
The Belinski-Khalatnikov-Lifshitz scenario concerns the existence of  generic singularity of \linebreak
general  relativity. At the singularity, there is a breakdown of all  known laws of physics.
Quantization of this scenario leads, however, to regular quantum evolution. The singularity
is avoided by a quantum bounce. It is fairly probable that quantum general relativity, to be \linebreak
constructed, would be free from singularities. Thus, it could be used to address issues such as the
quantum fates of cosmological and black holes singularities.
\end{abstract}


\maketitle

\section{Introduction}

Based on the assumption that the universe is spatially isotropic and homogeneous, Alexander Friedmann in 1922 derived
simple dynamics from Einstein's field equations. The solution to this dynamics includes gravitational singularity.
It is characterised by the divergence of gravitational and matter field invariants so that there is breakdown of all
the known laws of physics at the singularity. Friedmann's-type model, called Friedmann-Lema\^{i}tre-Robertson-Walker
universe, is commonly used in cosmology and astrophysics. However, in 1946, Evgeny Lifshitz found that the isotropy
is unstable in the evolution towards the singularity \cite{EL}. This important discovery initiated extensive examination
of the dynamics of anisotropic but homogeneous models, that is, Bianchi-type, in particular the Bianchi VIII and IX
 \cite{Belinski:2014kba}. The result of these investigations carried out by Belinski, Khalatnikov and Lifshitz (BKL),
 led to the conclusion that general relativity includes the generic solution with the singularity \cite{BKL}. Roughly
 speaking, by generic solution one means that it corresponds to non-zero measure subset of all initial data, it is
 stable against perturbation of the initial data, and depends on some arbitrary functions of space. This conjecture
 concerns  both cosmological and astrophysical singularities.  The BKL scenario can be seen in the low energy bosonic
 sectors of all five types of superstring models \cite{TMH}.

Quite independently, Roger Penrose \cite{RP} proved, among other things,   that under some conditions spacetime may
include incomplete geodesics. They are called singular despite they do not need to imply that the invariants diverge.
This theorem states little about the dynamics of the gravitational field near the end points of such pathological geodesics.
Thus, it is of little usefulness in the context of finding possible quantum dynamics.

On the contrary, the BKL assumption states that the terms with temporal derivatives in the dynamics dominate over
the terms with spatial derivatives when approaching the singularity. Consequently, the points in space decouple
and the dynamics becomes effectively the same as of the non-diagonal (general) Bianchi IX universe (see App. A for
more details). The dynamics
of the latter towards the singularity includes infinite number of oscillations of gravitational field, which lead
to a chaotic process that finally approaches, in finite proper time, the singularity. Described evolution is
called the BKL scenario.

\section{Classical dynamics}

The asymptotic form of the BKL scenario can be derived from the
general forms of the dynamics of the Bianchi IX spacetime if one makes the following assumptions \cite{book}:
(i) stress-energy tensor components can be ignored, (ii) Ricci tensor components $R^0_a$  have negligible influence
on the dynamics, and (iii) anisotropy of space may grow without bound. These assumptions  lead to enormous
simplification of the mathematical form of the dynamics. It can be well approximated by the following system
of equations \cite{bkl,book}:
\begin{equation}\label{L1}
\frac{d^2 \ln a  }{d \tau^2} = \frac{b}{a}- a^2,~~~~\frac{d^2 \ln b
}{d \tau^2} = a^2 - \frac{b}{a} + \frac{c}{b},~~~~\frac{d^2 \ln c
}{d \tau^2} = a^2 - \frac{c}{b},
\end{equation}
\begin{equation}\label{L2}
\frac{d\ln a}{d\tau}\;\frac{d\ln b}{d\tau} + \frac{d\ln
a}{d\tau}\;\frac{d\ln c}{d\tau} + \frac{d\ln b}{d\tau}\;\frac{d\ln
c}{d\tau} = a^2 + \frac{b}{a} + \frac{c}{b} \, ,
\end{equation}
where $\,a=a(\tau),\, b=b(\tau)$ and $\,c=c(\tau)$ are called the directional scale factors.

Equations \eqref{L1} and \eqref{L2} define a highly nonlinear coupled system of equations. They present the
essence of the BKL scenario. By applying dynamical system analysis to the above evolutions, it can be observed that
it has critical points of non-hyperbolic type so that the dynamics cannot be approximated by linearised equations \cite{EW}.
The space of the critical points is related to the gravitational singularity. Numerical simulations of the evolution of the
non-diagonal BIX model towards the singularity confirm the existence of the asymptotic form of the dynamics \cite{CNW}.

The evolution defined by Eqs. \!\eqref{L1}--\eqref{L2} is different from the commonly known Misner's mixmaster
dynamics \cite{Mis} specific to the diagonal (simplified) Bianchi IX model \cite{ewa}. The mixmaster dynamics has different
symmetry and does not take into account the effect of rotation of the so-called Kasner's axes. Other important difference
is that the system \eqref{L1}--\eqref{L2} describes the asymptotic dynamics of both BVIII and BIX spacetimes \cite{book}.

The existence of the generic singularity in solutions to Einstein's equations signals the existence of the limit
of validity of general relativity and means that this classical theory is incomplete. It is expected that the
imposition of quantum rules onto general relativity may lead to quantum theory devoid of singularities to be used to explain
observational  data.  Thus, it is tempting to quantize the dynamics defined by Eqs. \!\eqref{L1}--\eqref{L2} to determine if
one can get regular quantum dynamics.

The first step towards quantization of the asymptotic dynamics is  rewriting it in terms of the Hamiltonian system
devoid of any dynamical constraint. Making use of the reduced phase space technique enables rewriting the dynamics
\eqref{L1}--\eqref{L2} in the form of the  Hamiltonian system \cite{EW}:
\begin{eqnarray}
 \label{b1} d q_1 / d t &=& \partial H / \partial p_1 =  ( p_2 - p_1 + t ) / 2 F,\\
 \label{b2} d q_2 / d t &=& \partial H / \partial p_2 =  ( p_1 - p_2 + t / 2 F,\\
\label{b3} d p_1 / d t &=& - \partial H / \partial q_1 = (2 e^{2 q_1}- e^{q_2 - q_1}/ F  ,\\
\label{b4} d p_2 / d t &=&- \partial H / \partial q_2 = - 1 + e^{q_2 - q_1} / F,
\end{eqnarray}
where the Hamiltonian reads
\begin{equation}\label{ham1}
H(q_1, q_2; p_1, p_2; t) := - q_2 - \ln F(q_1,q_2,p_1,p_2, t),
\end{equation}
and where
\begin{equation}\label{b5}
F:= -e^{2 q_1}-e^{q_2 - q_1}-\frac{1}{4}(p_1^2 + p_2^2 +t^2)+\frac{1}{2}(p_1 p_2 + p_1 t + p_2 t)> 0.
\end{equation}
The the canonical variables $\{q_1,q_2,p_1,p_2\}$ and the evolution parameter $t$ are related to the
scale factors $\{a, b, c\}$.

The singularity of this dynamics turns out to be  defined by the condition:
\begin{equation}\label{qw8}
  q_1 \rightarrow - \infty,~~~q_2 - q_1 \rightarrow -\infty,~~~F\rightarrow 0^+~~~ \mbox{as}~~~t \rightarrow 0^+.
\end{equation}

The Hamiltonian \eqref{ham1} is not of polynomial-type so that canonical quantization cannot be applied.
Since the physical phase space consists of the two half-planes, it can be identified with the Cartesian
product of two affine groups. Thus, quantization can be carried out using the affine coherent states (ACS)
method \cite{AWG,AW}.

\section{Quantum dynamics}

Here we outline the main features  of the affine coherent states quantization scheme. More details are given
in App. B.

Roughly speaking, by  quantization of classical system represented  by observables defined on phase space,
we mean\footnote{We specify the simplest case of quantization. }:
\begin{itemize}
  \item ascribing to that system   self-adjoint operators acting in   Hilbert space ,
  \item ascribing to  Hamilton's dynamics  Schr\"{o}dinger's dynamics ,
  \item examination of  time dependance of  probability amplitude .
\end{itemize}

\subsection{Choice of Hilbert space}

The  physical phase space $\Pi$ consists of the two  half-planes:

\[ \Pi = \Pi_1 \times \Pi_2 := \{(q_1, p_1) \in \dR \times \dR_+\} \times  \{(q_2, p_2) \in \dR \times \dR_+\} ,\]
where $\dR_+ := \{x \in \dR~|~x>0 \}$.

Separately, $\Pi_1$ and $\Pi_2$ can be identified with the affine group Aff$(\dR)$.
This group has  the unitary irreducible representation  realized in the Hilbert space $ L^2(\dR_+, d\nu(x))$, where  $d\nu(x)=dx/x$,   defined by
\[  U(q,p)\psi(x)= e^{i q x} \psi(px).\]
This enables defining the  continuous family of  affine coherent states  $|q,p\rangle \in L^2(\dR_+, d\nu(x))$ as follows
\begin{equation}\label{im2}
 |q,p\rangle = U(q,p)|\phi\rangle \; ,
\end{equation}
where $|\phi\rangle \in L^2(\dR_+, d\nu(x))$, is the so-called   fiducial  vector,
which is a free  `parameter' of this quantization scheme.

\subsection{Quantum observables}

The irreducibility of the representation leads (due to Schur' lemma)  to the  resolution of the unity in $L^2(\dR_+, d\nu(x))$:
\begin{equation}\label{im44}
 \frac{1}{A_\phi} \int_{\Pi} d\mu (q,p)  |q,p\rangle \langle q,p| = \id \; ,
\end{equation}
where $d\mu (q,p) := dq\; dp/p^2$ is the left invariant measure on $\Pi$,  and where
$A_\phi := \int_0^\infty |\phi(x)|^2 \,\frac{dx}{x^2}< \infty$ is a constant.

Using \eqref{im44}, enables  quantization of any observable $f: \Pi \rightarrow \dR $ as follows:

\begin{equation}\label{im8}
   f \longrightarrow \hat{f} = \frac{1}{ A_\phi} \int_{\Pi} d\mu (q,p) |q,p\rangle f(q,p) \langle q,p| \, .
\end{equation}

The operator $\hat{f}$ is   symmetric (Hermitian) by construction.  No ordering  ambiguity occurs
(familiar disaster of canonical quantization).

\subsection{Quantum BKL scenario}

Since the Hamiltonian \eqref{ham1} is the generator of the evolution in the physical phase space, the Hermitian
operator corresponding to it can be used to define the
Schr\"{o}dinger equation (units are chosen so that  $~\hbar = 1 = c = G$).
For the case \eqref{ham1}, the following equation is obtained \cite{AWG}:
\begin{equation}\label{ns9}
i \frac{\partial}{\partial t} \Psi(t,x_1,x_2) =
 \left( i \frac{\partial }{\partial x_2}  -\frac{i}{2x_2} - K(t,x_1,x_2)
 \right) \Psi(t,x_1,x_2)\, ,
\end{equation}
where
\begin{equation} \label{ns6}
K = \frac{1}{A_{\Phi_1} A_{\Phi_2}}\;  \int_0^\infty  \frac{dp_1}{p_1^2} \int_0^\infty  \frac{dp_2}{p_2^2}\,
\ln\big(F_0(t, \frac{p_1}{x_1},\frac{p_2}{x_2})\big)
|\Phi_1(x_1/p_1 )|^2 |\Phi_2(x_2/p_2 )|^2 \, ,
\end{equation}
and where
\begin{equation}\label{ns1}
 F_0 (t, p_1,p_2) := p_1 p_2-\frac{1}{4} (t-p_1-p_2)^2 \, .
\end{equation}

The  general solution to the  Schr\"{o}dinger equation \eqref{ns9} reads
\begin{equation}\label{ns10}
\Psi=\eta(x_1,x_2+t-t_0)\, \sqrt{\frac{x_2}{x_2+t-t_0}}
\, \exp\left(i \int_{t_0}^t K(t',x_1,x_2+t-t')\,dt' \right) \, ,
\end{equation}
where $ t \ge t_0 >0$, and where $\eta (x_1, x_2):= \Psi (t_0, x_1,x_2)$ is the
initial state satisfying the condition
\begin{equation}\label{nor2}
 \eta(x_1,x_2) = 0~~~~\mbox{for}~~~~x_2 < t_H \, ,
\end{equation}
with $t_H > 0$ being the parameter of our model.

For $t < t_H$ we get
\begin{equation} \label{nor3}
\BraKet{\Psi(t)}{\Psi(t)} =
\int_0^\infty \frac{dx_1}{x_1}  \int_{t_H}^\infty \frac{d{x_2}}{x_2}\,
|\eta(x_1,x_2)|^2 \,,
\end{equation}
so that the  inner product is time independent, which implies that
the  quantum evolution is  unitary for $t > 0$.

The operator of the  time reversal, $\hat{T}: \mathcal{H} \rightarrow \mathcal{H} $, is defined to be
\begin{equation}\label{qb1}
   \hat{T}\, \psi (t, x_1, x_2) = \tilde{\psi} (t, x_1, x_2) :=  \psi (- t, x_1, x_2)^\ast,~~~\mbox{where}~~~\psi \in \mathcal{H} \, .
\end{equation}

Due to \eqref{ns9}, the  Schr\"{o}dinger equation for  $\tilde{\psi}$ reads
\begin{equation}\label{qb2}
i \frac{\partial}{\partial t} \tilde{\Psi}(t,x_1,x_2) =
 \left(- i \frac{\partial }{\partial x_2}  +\frac{i}{2x_2} - K(-t,t,x_1,x_2)
 \right) \tilde{\Psi}(t,x_1,x_2)\, .
\end{equation}

The general  solution to \eqref{qb2}, for $t < 0$,  is found to be
\begin{equation}\label{qb3}
\tilde{\Psi}=\eta(x_1,x_2 + |t| - |t_0|)\, \sqrt{\frac{x_2}{x_2 + |t| - |t_0|}}
\, \exp\left(i \int_{t_0}^t K(-t',x_1,x_2-t+t')\,dt' \right) \, ,
\end{equation}
where $ |t|\geq |t_0|$, and where  $\eta (x_1, x_2):= \tilde{\Psi} (t_0, x_1,x_2)$ is the initial state.

The  unitarity of the evolution (with $t_0 = 0$) can be obtained again if
\begin{equation}\label{qb4}
 \eta(x_1,x_2) = 0~~~~\mbox{for}~~~~x_2 < |t_H| \, ,
\end{equation}
which corresponds to the condition \eqref{nor2}.

Since the solutions \eqref{ns10} and \eqref{qb3} differ only by the
corresponding phases, the probability density is  continuous at $t=0$, which
means that we are dealing with  quantum bounce at $t=0$ (that marks the classical
singularity). This result is robust in the sense that it does not depend on the details of the ACS quantization \cite{AW}.

Equations \eqref{L1} and \eqref{L2} define the best prototype of the BKL scenario \cite{Belinski:2014kba,bkl}.
The quantum version of it presents a unitary evolution. The generic gravitational singularity
is avoided by a quantum bounce. This  result strongly suggests that quantum general relativity
has a good chance to be free from singularities so that  preliminary versions of it can be applied to address the issues
of cosmological and  black hole singularities.

\section{Prospects}

Cosmological model can be used to describe  black hole, BH,   after  imposing the condition that one deals with
an  isolated object.  This may be reduced to the problem of  merging finite region  of specific
spacetime with the Schwarzschild  spacetime \cite{WI}. The results concerning Hamiltonian formulation of dust cloud
collapse are promising \cite{Kwidzinski:2020xyd}. The quantization of the  Oppenheimer-Snyder model,  done recently
within affine coherent states quantization, reveals quantum bouncing scenario \cite{WT}. One may expect
that quantization of the collapsing star modelled by the Lema\^{i}tre-Tolman-Bondi spacetime, that may include naked or
covered singularity, can bring highly interesting results.

The sophisticated approach would be modelling of an isolated compact object with the Bianchi IX spacetime. Here,
the merging process is an open issue. Near the singularity, a link with the BKL scenario is expected. Since BIX
dynamics has strong anisotropic oscillatory modes, it is expected that BIX BH would radiate gravitational waves so
that one might detect them. The major challenge is however the construction of the rotating BH which might be used
in the description of the real black holes.

Quantum bounce, i.e. black to white hole transition,  may lead to astrophysical  small bang (analogy with
cosmological  Big Bang). Quantum gravity may be used to get  insight into the origin of numerous highly
energetic  explosions in distant galaxies like  GRBs, pulsars, etc, and  vice versa.


\appendix

\section{Metric of the Bianchi IX spacetime}

The  general form of a line element of the Bianchi IX model, in the
synchronous reference system, reads:
\begin{equation}\label{d1}
ds^2 = dt^2 - \gamma_{ab}(t)e^a_\alpha e^b_\beta dx^\alpha
dx^\beta ,
\end{equation}
where $a,b,\ldots$ run from $1$ to $3$ and label frame vectors;
$\alpha,\beta,\ldots$ take values $1,2,3$ and concern space
coordinates, and where $\gamma_{ab}$ is a spatial metric.

The  homogeneity of the Bianchi IX model means that the
three independent differential 1-forms $e^a_\alpha dx^\alpha$ are
invariant under the transformations of the  isometry group of the
Bianchi IX model.

The cosmological  time variable $t$ is redefined as
follows:
\begin{equation}\label{d2}
    dt = \sqrt{\gamma}\; d\tau,~~~~~\gamma := det[\gamma_{ab}]
\end{equation}
where $\gamma$ is the  volume density, and $ \gamma
\rightarrow 0$ denotes the singularity.

\section{Affine coherent states quantization}

The phase space $\Pi = \dR \times \dR_+$ can be  identified with the affine group $G = Aff(\dR)$
by defining the multiplication law as follows
\begin{equation}\label{c1b}
(q^\prime, p^\prime)\cdot (q, p) = (p^\prime q+ q^\prime, p^\prime p ),
\end{equation}
with the unity $(0,1)$ and the inverse
\begin{equation}\label{c2b}
(q^\prime, p^\prime)^{-1} = (-\frac{q^\prime}{p^\prime}, \frac{1}{p^\prime}).
\end{equation}

The affine group has two, nontrivial, inequivalent  irreducible unitary
representations.  Both are realized in the Hilbert space
$\mathcal{H}=L^2(\dR_+, d\nu(x))$, where $d\nu(x)=dx/x$ is the invariant
measure on the multiplicative group $(\dR_+,\cdot)$.
In what follows we choose the one defined by
\begin{equation}\label{im1b}
U(q,p)\psi(x)= e^{i q x} \psi(px)\, .
\end{equation}
The integration over the affine group reads
\begin{equation}\label{m5}
  \int_G d\mu (p,q) := \int_{-\infty}^\infty dp \int_0^\infty dq /q^2\, ,
\end{equation}
where the measure $d\mu (p,q)$ is left invariant.

Fixing the normalized vector
$\Ket{\Phi} \in L^2(\dR_+, d\nu(x))$, called the  fiducial vector, one can define
a continuous family of  affine coherent states $\Ket{q,p} \in L^2(\dR_+, d\nu(x))$
as follows
\begin{equation}\label{im2}
\Ket{q,p} = U(q,p) \Ket{\Phi}.
\end{equation}
The  irreducibility of the representation, used to define the coherent
states \eqref{im2}, enables making use of Schur's lemma, which leads
to the  resolution of the unity in $L^2(\dR_+, d\nu(x))$:
\begin{equation}\label{im4}
\int_{G}  d\mu(q,p) \Ket{q,p}\Bra{q,p} = A_\Phi\;\id \; ,
\end{equation}
where
\begin{equation}\label{im3b}
A_\Phi = \int_0^\infty \frac{dp}{p^2} |\Phi(p)|^2 \, .
\end{equation}

Making use of the resolution of the unity \eqref{im4}, we define the
quantization of a  classical observable $f: \Pi \rightarrow \dR$ as follows
\begin{equation}\label{im8}
 f \longrightarrow  \hat{f} :=
\frac{1}{A_\Phi}\int_{\Group{G}} d\mu(q,p)
\Ket{q,p} f(q,p) \Bra{q,p}  \, ,
\end{equation}
where $\hat{f}: \mathcal{H} \rightarrow \mathcal{H}$ is the corresponding
quantum observable.
The mapping \eqref{im8} is  covariant in the sense that one has
\begin{equation}\label{cov}
  U(\xi_0) \hat{f} U^\dag (\xi_0) =
\frac{1}{A_\Phi}\int_{\Group{G}} d\mu_L(\xi)
\Ket{\xi} f(\xi_0^{-1}\cdot \xi) \Bra{\xi} =  \widehat{\mathcal{L}^L_{\xi_0}f}   \, ,
\end{equation}
where $\mathcal{L}^L_{\xi_0} f(\xi) = f(\xi_0^{-1}\cdot \xi)$ is the left shift operation,
and where $\xi_0^{-1}\cdot \xi= (q_0,p_0)^{-1}\cdot (q,p) = (\frac{q-q_0}{p_0},\frac{p}{p_0})$,
with $\xi := (q,p)$.  It means,  no point in the phase space $\Pi$   is privileged.

Eq. \!\eqref{im8} defines a linear mapping and the observable $\hat{f}$ is a
 symmetric  operator by the construction. Let us evaluate the norm of this operator:
\begin{equation}
\label{OperNormQuantOper}
\Vert\hat{f}\Vert \leq
\frac{1}{A_\Phi}\int_{\Group{G}} d\mu (q,p)
|f(q,p)|  \Vert \Ket{q,p}\Bra{q,p} =
 \frac{1}{A_\Phi}\int_{\Group{G}} d\mu(q,p) |f(q,p)| \, .
\end{equation}
This implies that, if the classical function $f$ belongs to the space of
integrable functions $L^1(\Group{G}, d\mu_L(q,p))$, the operator
$\hat{f}$ is  bounded so that it is a  self-adjoint operator.  Otherwise, it is
defined on a dense subspace of $L^2(\dR_+, d\nu(x))$ and its possible
self-adjointness becomes an open problem to be examined.


\begin{thebibliography}{99}

\bibitem{EL} E. M. Lifshitz, J. Phys. (USSR) {\bf 10}, 116 (1946); E. M. Lifshitz and I. M. Khalatnikov, Adv. Phys. {\bf 12}, 185 (1963).

\bibitem{Belinski:2014kba} V.~A.~Belinski,
  Int.\ J.\ Mod.\ Phys.\ D {\bf 23}, 1430016 (2014).

\bibitem{BKL}
V. A. Belinskii, I. M. Khalatnikov and E. M. Lifshitz, Adv. Phys.
{\bf 19},  525 (1970); {\bf 31}, 639 (1982).

\bibitem{TMH} T. Damour, M. Henneaux, and H. Nicolai, Class. Quant. Grav. {\bf 20} (2003) R145.

\bibitem{RP} R. Penrose, Phys. Rev. Lett. {\bf 14}, 57 (1965).

 \bibitem{bkl} V. A. Belinskii, I. M. Khalatnikov, and M. P. Ryan,
``The oscillatory regime near the singularity in Bianchi-type IX
universes'',  Preprint {\bf 469} (1971), Landau Institute for
Theoretical Physics, Moscow (unpublished); the work due to V. A. Belinskii and I. M. Khalatnikov
is published as sections 1 and 2   in  M. P. Ryan, Ann. Phys. {\bf 70},  301 (1971).

\bibitem{book} V.~Belinski and M.~Henneaux,
{\em The Cosmological Singularity} (Cambridge University Press, Cambridge, 2017).

\bibitem{EW} E.~Czuchry and W. Piechocki, Phys.\  Rev.\  D {\bf 87},  084021 (2013).

\bibitem{CNW} C. Kiefer, N. Kwidzinski, and W. Piechocki, Eur. Phys. J.  C {\bf 78}, 691 (2018).

\bibitem{Mis} C. W. Misner, Phys. Rev. Lett. {\bf 22}, 1071 (1969); Phys. Rev. {\bf 186}, 1319 (1969).

\bibitem{ewa}
  E.~Czuchry, N.~Kwidzinski, and W.~Piechocki,
  Eur.\ Phys.\ J.\ C {\bf 79}, 172 (2019).



\bibitem{AWG} A. G\'{o}\'{z}d\'{z}, W. Piechocki, and G. Plewa, Eur. Phys. J. C {\bf 79}, 45 (2019).

\bibitem{AW} A. G\'{o}\'{z}d\'{z} and W. Piechocki, Eur. Phys. J. C {\bf 80}, 142 (2020).

\bibitem{WI} W. Israel, Nuovo Cimento B {\bf 44}, 1 (1966); {\bf 48}, 463 (1966).

\bibitem{Kwidzinski:2020xyd}
  N.~Kwidzinski, D.~Malafarina, J.~Ostrowski, W.~Piechocki, and T.~Schmitz,
 Phys.\  Rev.\  D {\bf 101}, 104017   (2020).

\bibitem{WT} W. Piechocki and T. Schmitz, arXiv:2004.02939.

\end{thebibliography}
\end{document}